# Novel Approach to the Dark Matter Problem: Primordial Intermediate-Mass Black Holes

Paul H. Frampton

Introduction

Fifty years ago, particle phenomenology and theoretical cosmology were quite separate areas of study, and particle theorists at that time treated cosmology with some condescension because unlike for particles the uncertainties in the data were very large, often a factor of two or more. Presently this has completely changed because cosmology has become a precision subject where many measurements have been made at the fraction of a percent level.

For example, the so-called visible universe (visible by electromagnetic radiation) is a sphere centered e.g. at the Earth, or equivalently at any other point, with mass and radius now measured as accurately as could be those of a basketball measured by a ruler and a bathroom scale. This precision, and the explosion of new experimental data, is in part what has attracted many former particle phenomenologists to turn their minds to the study of cosmology.

One of the biggest puzzles in today's cosmology theory is the nature of dark matter which makes up about one quarter of the energy content of the universe. The latest data suggest that the energy make-up is 68.4% dark energy, 26.7% dark matter and 4.9% normal matter. We shall not address dark energy here, only dark matter. Of the matter, only about one sixth is normal matter while the other five sixths is matter detected only by gravitational interaction and inaccessible to electromagnetic radiation.

Dark matter was first discovered in the Coma cluster by Fritz Zwicky in 1933, and its presence in many, and presumably most, galaxies was confirmed by the study of rotation curves in the 1970s especially by Vera Rubin and her group. It is now beyond question that the dark matter is there, and the Bullet Cluster collision confirms its presence directly.

It has been a bit like a detective novel to determine what are the constituents of the dark matter. In the tradition of Agatha Christie a couple of very likely suspects have shown up and seemed very likely but as time passed these strong suspects became less probable only to be replaced now by a quite different and perhaps more probable one. We shall describe how the confluence of several lines of argument suggest that dark matter may not be comprised of microscopic particles but be purely astrophysical in the form of macroscopic black holes.

The standard model of fundamental interactions does not contain a suitable candidate for the dark matter constituent, but well motivated extensions beyond the standard model do. Therefore it is important that we shall first discuss these, the most popular of which are the axion and the WIMP.

Particle theory

For any particle phenomenologist, it is irresistible to invent a theory extending the standard model to one which contains a a suitable candidate to be a dark matter constituent. This assumes that the appropriate dark matter constituent is a microscopic elementary particle, an assumption which we shall re-examine and query later.

Two excellent examples of particle theory models exist motivated by physics totally unrelated to dark matter, and each provides an attractive candidate for the dark matter constituent.

The first such model is motivated by solution of the so-called strong CP problem. Part of the standard model is quantum chromodynamics (QCD) as the theory of strong interactions, of quarks interacting via gluon exchange. Its lagrangian is very similar to that for quantum electrodynamics (QED) so at first sight it appears to respect CP symmetry. However, because of the existence of classical field configurations called instantons, which have no analogue in QED, QCD does not naturally conserve CP. Instead a QCD parameter must be fine-tuned to be less than $10^{-9}$ to agree with the experimental upper limit on CP violation.

One attractive possibility is to introduce a second scalar doublet and a global Peccei-Quinn U(1) symmetry which naturally allows the relevant QCD parameter to relax to zero and hence to restore CP symmetry. However, this global symmetry is spontaneously broken, giving rise to a light pseudoscalar called the axion. In its original version the axion mass was a few times 100keV but this was soon excluded experimentally. The axion model was then changed so that the axion became very much lighter and very much more weakly coupled, hence the name invisible axion. This invisible axion is the proposed candidate for a dark matter constituent and its mass is between a micro eV and a milli eV. The manipulation to render the axion invisble may already cause suspicion but more serious is the fact that when the invisible axion model is coupled to gravity in the simplest way, a fine tuning of a new parameter to one part in $10^{-40}$ is necessary. This can be avoided by further elaborations of the model. The invisible axion may exist but it is suspiciously unnatural.

A second generalization of the standard model is motivated by the observation that the mass of the Higgs boson has a quadratic divergence which drives it to a very high value and fine-tuning is necessary to keep it at its experimental value of only 125 GeV. Extending the standard model by adding the, so far experimentally unobserved, property of supersymmetry gives a model in which the aforementioned quadratic divergence is absent. There is a precise cancelation between the fermionic and bosonic contributions. Supersymmetry relates bosons and fermions and to each particle of the standard model is assigned a superpartner with spin differing by one half. A symmetry called R parity is necessary to avoid too-fast proton decay. The normal particles have R=+1 and superpartners have R=-1. Consequently the lightest R=-1 particle is stable and a candidate for dark matter.

This lightest superpartner is most commonly taken to be the neutralino which is a superposition of superpartners of the Z, $\gamma$ and H normal particles. If supersymmetry is broken at the weak scale, about 100 GeV, as necessary to stabilize the Higgs mass, then the neutralino mass will be near the weak scale too. This particle necessarily has weak interactions just as the W and H normal particles do and so is called a WIMP for Weakly Interacting Massive Particle. This acronym WIMP, along with the axion, were appropriated by the cosmology community as the most popular dark matter candidates for at least the last twenty years.

Somewhere around 1990 this WIMP developed a life of its own as a dark matter candidate and millions of euros were, and still are being, spent searching for WIMPs. The historical link to weak-scale supersymmetry was widely forgotten. Experiments at the Large Hadron Collider (LHC), however, in 2014 and 2015 disfavored the existence of weak-scale supersymmetry because no superpartners were found. Consequently the WIMP is also disfavored. Both of the prime dark matter candidates arising from extending the standard model of particle theory, the axion and the WIMP, are in enough trouble that we should look elsewhere.

Baryon number
Baryon number is defined to be B=+1 for nucleons, B=-1 for antinucleons and B=0 for leptons and bosons. No violation of B has ever been observed so it seems as rigorously conserved as electric charge. For reasons to become clear, baryon number conservation plays an important role in our identification of the dark matter. Having failed to find acceptable microscopic dark matter constituents, we consider Massive Compact Halo Objects or MACHOs. MACHOs are unlike WIMPs in that they cannot be detected in terrestrial experiments, only by looking up *i.e.* with purely astronomical observations.

In the early universe starting about one second and ending about one hundred seconds after the big bang is the period of Big Bang Nucleosynthesis (BBN), a period that is very well understood. The temperature T varies as $t^{-1/2}$ where t is cosmic time. Actually $t T^2$ ~ 1MeV so the temperature drops from about 1MeV to about 100keV during BBN. Helium is produced by binding protons and neutrons giving close to 25% by mass of helium in good agreement with present data. No element heavier than lithium appears. The BBN calculation was first computed by Ralph Alpher in 1948. The point is that as a result, the baryon content of the universe is constrained to comprise at most one sixth of the total matter. It is thus impossible, given the BBN constraint, that baryonic material make up all, or even a majority of, the dark matter.

Examples of MACHOs are compact non-luminous objects such as white dwarfs, neutron stars, brown dwarfs and unassociated planets. All of these are baryonic. Then, there are black holes but if these arise from gravitational collapse of stars which are baryonic they are similarly restricted by the BBN constraint.

This suggests that, if it is possible to form them, Primordial Black Holes (PBHs) which are made during the radiation dominated era are the dark matter MACHO candidate of choice.

Entropy

Another idea which plays an important role in identifying the dark matter is the entropy of the universe. Entropy was a concept within thermodynamics introduced in the 19th century by Clausius in 1865 and whose significance in statistical physics was famously analyzed by Boltzmann by his H theorem in 1871. This led to a statistical understanding of the second law of thermodynamics, that for an isolated system the entropy can only increase. The result puts Boltzmann at the top of theory in the 19th century along with Maxwell. Perhaps Boltzmann ranks the higher because, although we understand Maxwell's electromagnetic equations of 1865, a full understanding of Boltzmann's 1871 theorem can even now be exasperatingly elusive.

Black holes have the maximum entropy possible for a given volume, and the entropy is curiously proportional to the surface area. Given a black hole with mass $M_{BH}=\eta M_{SUN}$, its dimensionless entropy is given by $S_{BH} \sim 10^{77} \eta^2$. Thus, a Supermassive Black Hole (SMBH) with $\eta=10^7$ has an entropy $10^{91}$ which is larger than the entropy ($\sim 10^{88}$) of all the Cosmic Microwave Background (CMB) in the visible universe. This gives an indication of by how much the black holes concentrate entropy and dominate the entropy of the universe. If we take $10^{11}$ galaxies each with a $10^7 M_{SUN}$ mass SMBH then the entropy of the interior of the visible universe from SMBHs alone is $S \sim 10^{102}$.

One can reasonably expect the interior entropy to be more, and the only way that can happen is if there are far more black holes, for example trillions and trillions of PIMBHs acting as the dark matter in the galaxies and clusters of galaxies.
It must be emphasized that this is only a *suggestive* argument based on the second law of thermodynamics and treating the universe as an isolated system. whether the second law is respected in this way is a dynamical question requiring numerical simulations far beyond anything practicable because of their complexity. Nevertheless, the entropy of the universe argument is an indicator. Just to give one example, if the dark matter PIMBHs all have mass $10^4 M_{SUN}$ and there are $10^8$ of them in each of $10^{11}$ galaxies their entropy is $S \sim 10^{104}$ which is two orders of magnitude more than for the SMBHs.

Primordiality

There are two types of black hole, those which are formed by gravitational collapse of a baryonic object and those which are formed primordially *i.e.* during the radiation-dominated era near the beginning of the expansion. From our previous discussion of baryon number B, the majority of the dark matter black holes must be primordial to avoid the BBN constraint. Let us then focus on the dark matter known in the halo of a galaxy, for example the Milky Way which includes our Sun. MACHO searches have ruled out masses below about $10 M_{SUN}$ for making all the dark matter so that is a lower mass limit. There is also an important upper limit on the halo MACHO mass arising from stability of the disk dynamics, a limit which is about $10^6 M_{SUN}$.
These two bounds then leave open the so-called intermediate-mass range $10 M_{SUN} < M_{PIMBH} < 10^5 M_{SUN}$, defining precisely what we meant by Primordial Intermediate-Mass Black Hole (PIMBH).

In order to form such PIMBHs in the early universe, two requirements are (i) high density and (ii) strong inhomogeneity. The first requirement is straightforward in the early universe where the density can be arbitrarily high. The second requirement is more non-trivial but it has been shown that in one version of inflation, named hybrid inflation, large enough variation in the density is possible to form PBHs of very high mass. This arises because of a parametric resonance between the inflaton field which drives a first stage of inflation and the so-called waterfall field which drives the second stage of inflation. In this model it is possible to form PBHs not only massive enough to be PIMBHs but with masses even up to $10^{17} M_{SUN}$, although such a gigantic black hole heavier than a cluster of galaxies seems extremely unlikely to exist in our universe. One peculiarity of the model is that the mass function of PBHs is sharply spiked at a particular mass value, implying all the PIMBHs have about the same mass. It is not yet understood whether this is an artifact of the specific model. In any case, the hybrid inflation model provides an existence theorem that PIMBHs can be formed primordially as required in this solution for dark matter.

Microlensing

Let us discuss how to search for the PIMBHs which we claim exist in millions, probably billions, within the Milky Way. There are two indirect methods which have been attempted, although the results are ambiguous. One indirect method is to study the behavior of wide binaries which are very weakly bound systems of two stars very far apart. The second is to look for distortion, either in spectrum or isotropy, of the CMB. The latter could arise from X-rays emitted by gas accreted on the MACHOs and downgraded to microwaves by Thomson scattering.

A far more direct method is to use microlensing where the MACHO acts as a gravitational lens which amplifies the intensity of a distant star as the MACHO transits the light path between the star and the Earth. This method was proposed by the late Bohdan Paczynski in 1986. Because good alignment between the MACHO lens and more distant star is crucial, it is necessary to scan millions of stars to find examples of microlensing events. The durations of such events depend on the MACHO mass, proportionately to the square root of the mass. The duration is about one year for a MACHO mass $25 M_{SUN}$, two years for $100 M_{SUN}$ and so on. In the 1990s, a MACHO collaboration working in Australia at Mount Stromlo Observatory successfully observed such microlensing events for masses up to about $20 M_{SUN}$. They, however, ended their project in about 2000 because they had found enough MACHOs only to make about 10% of the halo's dark matter and had no theoretical motivation to suspect heavier MACHOs. Higher duration microlensing events will be decisive about whether PIMBHs do form the dark matter. The recent detection of gravitational waves from a merger of two black holes can be interpreted as support for the theory.

Epilogue

If our discussion is correct, it provides a clear time-ordering for galaxy formation that the dark matter precedes star formation by half a billion years. Let us consider the history of the Milky Way. The constituents of the Milky Way's dark matter halo, PIMBHs, were produced in the era of radiation domination which ended at time t ~ 47 ky (red shift Z ~ 4760). Only much later, after 560 million years (Z ~ 8), did star formation begin in the Milky Way.

In this version of cosmic history, most of the large-scale structure formation including of galaxies such as the Milky Way progresses during the half billion years represented by the red shifts 4760 > Z > 8. This stage importantly involves *only* dark matter. Baryonic astrophysical objects like the Sun and the Solar System appear only when Z < 8 and are secondary with respect to the Milky Way's formation. Assuming that we have identified correctly the constituents of the dark matter, our solution therefore sheds light on the history of large-scale structure formation.

Further reading
P.H. Frampton, *Searching for Dark Matter Constituents with Many Solar Masses*.
P.H. Frampton, *The Primordial Black Hole Mass Range.*
T. Axelrod, G. Chapline and P.H. Frampton*, Intermediate Mass MACHOS: a New Direction for Dark Matter Searches.*